\documentstyle[12pt]{article}

\if@twoside\oddsidemargin25mm
\else\oddsidemargin25mm\evensidemargin25mm\marginparwidth25mm\fi
\def\bl{\Big\{}                  \def\br{\Big\}}
\def\bpl{\Big(}                  \def\bpr{\Big)}
\def\bq{\begin{equation}}        \def\eq{\end{equation}}
\def\brr{\begin{eqnarray}}       \def\err{\end{eqnarray}}
\def\ba{\left(\begin{array}}     \def\ea{\end{array}\right)}
\def\der{\partial}
\def\rar{\rightarrow}

                \def\Oo{{\cal O}}
                \def\Aa{{\sl a}}

\def\Nn{{\cal N}}

\def\g{\gamma}                   \def\d{\delta}
\def\e{\epsilon}                 
                   
\def\ve{\varepsilon}             
\def\a{\alpha}                   
\def\vphi{\varphi}               
\def\G{\Gamma}

\def\cN{{\cal N}}
\def\pr{{\prime}}
\def\lam{\lambda}
\def\Lam{\Lambda}

\def\Ds{D \!\!\!\! /\,\,}

\def\tq3{\tilde{q}_3}
\def\bt{\bar{\tau}}

\def\xb{{\bar x}}                \def\tb{{\bar\tau}}

\newcommand{\nonu}{\nonumber \\ }

\newcommand{\dr}{\raise.3ex\hbox{$\stackrel{\leftarrow}{\partial }$}{}}
\newcommand{\dl}{\raise.3ex\hbox{$\stackrel{\rightarrow}{\partial}$}{}}
\newcommand{\eqn}[1]{(\ref{#1})}

\newsavebox{\uuunit}
\sbox{\uuunit}
     {\setlength{\unitlength}{0.825em}
     \begin{picture}(0.6,0.7)
        \thinlines
        \put(0,0){\line(1,0){0.5}}
        \put(0.15,0){\line(0,1){0.7}}
        \put(0.35,0){\line(0,1){0.8}}
        \multiput(0.3,0.8)(-0.04,-0.02){12}{\rule{0.5pt}{0.5pt}}
     \end {picture}}

\newsavebox{\zzzbar}
\sbox{\zzzbar}
  {\setlength{\unitlength}{0.9em}
  \begin{picture}(0.6,0.7)
  \thinlines
  \put(0,0){\line(1,0){0.6}}
  \put(0,0.75){\line(1,0){0.575}}
  \multiput(0,0)(0.0125,0.025){30}{\rule{0.3pt}{0.3pt}}
  \multiput(0.2,0)(0.0125,0.025){30}{\rule{0.3pt}{0.3pt}}
  \put(0,0.75){\line(0,-1){0.15}}
  \put(0.015,0.75){\line(0,-1){0.1}}
  \put(0.03,0.75){\line(0,-1){0.075}}
  \put(0.045,0.75){\line(0,-1){0.05}}
  \put(0.05,0.75){\line(0,-1){0.025}}
  \put(0.6,0){\line(0,1){0.15}}
  \put(0.585,0){\line(0,1){0.1}}
  \put(0.57,0){\line(0,1){0.075}}
  \put(0.555,0){\line(0,1){0.05}}
  \put(0.55,0){\line(0,1){0.025}}
  \end{picture}}

\begin{document}

\begin{titlepage}

\begin{flushright}
ITP-SB-98-36 \\
KUL-TF-98/24 \\
hep-th/9805105 \\
\end{flushright}

\begin{center}
\vskip3em
{\large\bf $R$-Current Correlators in $\Nn=4$ super-Yang-Mills Theory
from anti-de Sitter Supergravity}

\vskip3em

{Gordon Chalmers$^{a}$\footnote{e-mail addresses: chalmers, hnastase, konrad,
siebelin@insti.physics.sunysb.edu}
Horatiu Nastase$^{a}$ \\
Koenraad Schalm$^{a}$,
Ruud Siebelink$^{ab}$\footnote{Post-doctoraal Onderzoeker FWO, Belgium}},

\vskip .9em

$^a$
{\it Institute for Theoretical Physics\\ State University of New York\\
 Stony Brook, NY 11794-3840, USA\\}

\vskip .9em

$^b$
{\it Instituut voor theoretische fysica, \\ Katholieke
Universiteit Leuven\\ B-3001 Leuven, Belgium\\}

\vskip .4cm
\end{center}

\vfill

\begin{abstract}
\noindent
We test the conjectured relationship between $\Nn=4$ super-Yang-Mills theory in four
dimensions and $IIB$ supergravity compactified on $\mbox{AdS}_5\times S_5$ by computing the
two- and three-point functions of $R$-symmetry currents.   We observe that the
integral expressions describing the general three-point  correlator on the 
supergravity side have a structure similar to one-loop triangle diagrams in 
$\cN=4$ super-Yang-Mills theory.  This allows us to compare the expressions on  both
sides of the $\mbox{AdS}$/CFT correspondence without the technical complications of the loop
integrations.  We confirm that the two- and three-point correspondence arises  at only
one-loop in the $\Nn=4$ super-Yang-Mills theory.   Higher-point functions as well as
further three-point functions may be analyzed similarly. 

\end{abstract}

\vfill

\end{titlepage}
\setcounter{footnote}{0}

\section{Introduction}

Recently there has been much progress in the understanding of conformal
field  theories in greater than two dimensions \cite{m1,ew,gkp}\footnote{See 
\cite{freedman} for a comprehensive list of recent work.}. These 
developments are based on the conjecture by Maldacena
that $d$-dimensional conformal field theories are dual to certain string
theories compactified to $d+1$-dimensional anti-de Sitter space \cite{m1}. In
\cite{ew,gkp} a prescription was given how to evaluate conformal field 
theory correlation functions starting
from string theory, or rather supergravity, on $\mbox{AdS}_{d+1}$; the boundary
values of the supergravity fields act as
sources for conformal operators. This scheme permits an evaluation of
superconformal correlators in the 't Hooft limit at strong effective 
coupling by examining the classical action, i.e. tree graphs, of IIB supergravity.  
There have been several recent tests of this conjecture involving the 
computation of correlators in \cite{muck,freedman,tsey}.   

In this paper we examine the two- and three-point function of $SU(4)$ $R$-currents in
$d=4$, $\cN=4$ super-Yang-Mills theory.  We compute these correlation functions from
supergravity on $\mbox{AdS}_5\times S_5$ and compare with the one-loop
$\cN=4$ super-Yang-Mills result. The fact that we make the comparison only at 
one-loop is surprising because the regime where the supergravity result is valid
coincides with  large 't Hooft coupling.   For the anomalous
$d_{abc}$ contribution this is clearly correct by the Adler-Bardeen theorem, but there
is no apparent reason why the same holds for the $f_{abc}$ term, save a possible
superconformal non-renormalization theorem as noted in 
\cite{freedman}. We will show that the supergravity result for both terms, if left as
an integral expression, has the structure of a field theory triangle diagram.  This
could be due to the  constrained kinematics of three-point functions as well as the
requirements of  conformal symmetry.  The one-loop calculations will be performed in
momentum space. This requires a regularization of the UV-divergences for which we have
chosen the method of dimensional reduction. Our procedure finds the correspondence
without having to evaluate all the integrals. This makes our method attractive for
further tests of the conjectured $\mbox{AdS}$/CFT correspondence involving higher-point
functions.

The full three-point correlator has a unique form dictated by conformal
invariance up to three undetermined constants, and their values are
determined solely from the OPE coefficients \cite{osborn1,osborn2}.  We reconstruct the
explicit position space form of the correlator from these OPE limits.  The individual
terms  occuring within the actual calculation of the full correlators generically
produce logarithms and dilogarithms. Conformal invariance requires that 
these are absent in the final expression for the correlator.  We show that 
this is the case.

The outline of this work is as follows.  In section 2 we briefly review the
prescription for the correspondence between the supergravity
theory and its superconformal cousin living on the boundary.
We find the $x$-space expression of the two-point function and
the OPE of the three-point function which we use to generate the
complete correlator expression.  In sections 3  and 4 we present the
calculation in momentum space  on both sides of the $\mbox{AdS}$/CFT correspondence
and verify that the two- and three-point functions match. This is done by
explicitly comparing the integral expressions.
In section 5 we present our conclusions.

\section{$\mbox{AdS}$/CFT Correspondence}

In this section we briefly discuss the characteristics of the
correspondence. On the one hand one considers type IIB string theory in an
$\mbox{AdS}_5 \times S_5$ background.
 If the $S_5$
carries $N$ units of five-form flux, the string theory is conjectured
\cite{m1} to be
dual to $\cN = 4$ $SU(N)$ super-Yang-Mills theory. The radius of the
five-sphere, which equals the radius of $\mbox{AdS}_5$, is given by
\bq R^2 = \a^\pr \sqrt{4\pi g_{st}N} \ , 
\eq
while the Yang-Mills coupling constant is given by
\bq g_{YM}^2 = g_{st} \ .
\eq
One may reliably approximate the full IIB string theory by
supergravity when $R^2/\a^{\prime}$ and $1/{g_{st}}$ are taken to be
large.
>From the Yang-Mills point of view, this corresponds to the 't
Hooft limit, $N \rar \infty$, in which the effective coupling,
\bq 
g^2_{\rm eff} = g^2_{YM} N \ , 
\eq
is kept fixed but large.  This means that one can study the super-Yang-Mills theory at
strong effective coupling, by considering tree diagrams of supergravity.

In \cite{ew,gkp} a prescription was given how to calculate
$\cN=4$ $SYM$ correlation functions from the IIB supergravity theory.
Sources for
conformal operators in the $SYM$-theory correspond to the boundary values of
the supergravity fields on $\mbox{AdS}_5 \times S_5$. The connected generating
functional for CFT-correlators is identified
with the supergravity action as a functional of the boundary values of the fields
\bq  
W[\phi_0] = \ln 
\langle e^{-J\cdot\vphi_0} \rangle_{CFT}=- S_{sugra, {AdS}}[\vphi[\vphi_0]] \ .
\label{cor}
\eq
However, in this relation there
are still several constants to fix: the relative
normalization $\lambda$ of the sources to the supergravity boundary values,
and an overall normalization $\omega$ of the action. More precisely
eq.~\eqn{cor} should read
\bq 
W[\phi_0] = \ln 
\langle e^{-J\cdot\vphi_0} \rangle_{CFT}=-\omega S_{sugra,
{AdS}}[\vphi[\lambda\vphi_0]] \ ,
\eq
where $\vphi[\lambda\vphi_0]$ are the fields in the bulk of $\mbox{AdS}$,
expressed through
their equations of motion as functions of the boundary values $\vphi|_{\der
{AdS}}=\lambda\vphi_0$.
For a generic connected correlation function this results in an overall factor of 
\brr
\langle J(1)J(2) \ldots J(n) \rangle_{CFT} 
= -\omega (-\lambda) ^n\frac{\delta ^n} 
 {\delta\vphi_0(1)...\delta\vphi_0(n)} S_{{AdS}}[\vphi_0] \Big|_{\vphi=0} \ .
\err  
In the following, the couplings for supergravity, $g^2$ and $k$, will correspond to
separate overall normalizations $\omega$.  We will reinstate boundary value to source 
normalizations $\lambda$ only at the end of our calculations of the 
correlators, when we make the comparison between $\cN=4$ super-Yang-Mills 
theory and $\mbox{AdS}$ supergravity.  

\subsection{Anti-de Sitter Supergravity}

The supergravity fields which correspond to $R$-symmetry
currents on the $SYM$-side are the $SU(4)$-gauge fields
$A_i$.  The action for the gauge fields
is independent of the details of the IIB compactification;
up to cubic terms the action including a Chern-Simons term is,

\brr 
S[A] &=& \frac{1}{2g^2} \int_{B_5} \d_{ab} \,
dA^a\wedge {}^*dA^b
 + f_{abc} \,dA^a \wedge {}^*
\{ A^b
     \wedge A^c\} \nonu
&& +\frac{ik}{32\pi^2} \int_{B_5} d_{abc} \, A^a \wedge
dA^b \wedge
     dA^c  \ .
\label{act}
\err 
Our $SU(4)$ Lie algebra conventions with anti-hermitian
generators $T_a$ are
\bq
T_aT_b = \frac{1}{2} \left( f_{ab}{}^c - i d_{ab}{}^c
\right) T_c \ .
\eq
The trace in a
representation $R$ is given by
$Tr_R (T_a T_b) = - C_R \d_{ab}$. As usual the
fundamental representation is normalized to
$C_f = \frac{1}{2}$.  

We parameterize the $\mbox{AdS}_5$ space by a set of coordinates
$x_0$, $\xb_i$ with the metric
\bq ds^2 = {1\over x_0^2} (dx_0^2 + d\xb^2) \ .
\eq
The $\mbox{AdS}$ radius has been set to unity.
The coordinate $x_0$ ranges from $0$ to infinity and the boundary of the
anti-de Sitter space is at $x_0=0$. Boundary indices $i=1, \ldots , 4$
are raised and lowered by $\d_{ij}$.

Following the prescription in \cite{ew,gkp} we can uniquely determine
the dependence of the
bulk $\mbox{AdS}_5$ gauge field $A_i^a$ on the
boundary values $\Aa_i^a$ through propagation with the
appropriate kernel in a certain gauge. Ghosts are not necessary
since we will only calculate tree diagrams. Using the
shorthand notation
$R(\xb_1;x_0,\xb) = x_0^2 + |\xb -\xb_1|^2$ the gauge field
and its exterior derivative in the bulk can be written
as\brr A^a (x_0,\xb) &=& d\xb_i  \int d^d \xb_1 \,\Aa_i^a
(\xb_1)
     \frac{x_0^{d-2}}{R(\xb_1;x_0,\xb)^{d-1}}  \nonu
     &&-  dx_0 \int d^d \xb_1\,  \Aa_i^a (\xb_1) \frac{1}{2(d-2)}
     \frac{\der}{\der\xb_1^i} \, \frac{x_0^{d-3}}{R(\xb_1;x_0,\xb)^{d-2}}
     \nonu
     dA^a (x_0,\xb)&=& d\xb_i \wedge d\xb_j  \int d^d \xb_1 \,
     \Aa_{i}^a (\xb_1) \frac{\der}{\der \xb_1^{j}}
     \frac{x_0^{d-2}}{R(\xb_1;x_0,\xb)^{d-1}}\nonu
     && - dx_0 \wedge d\xb_i \int d^d \xb_1 \,  \Aa_j^a (\xb_1)
     \frac{1}{2(d-2)} \left[ \d_{jk} \d_{li}- \d_{kl} \d_{ji}\right]
     \frac{\der}{\der \xb_1^k} \frac{\der}{\der \xb_1^l}
     \frac{x_0^{d-3}}{R(\xb_1;x_0,\xb)^{d-2}} \ . \nonu
\label{valueA}
\err
These expressions are to zeroeth order in the Yang-Mills couplings and
receive further
${\cal O} (\Aa^2)$ corrections due to interactions.

At the level of two- and three point functions these corrections 
do not play any role (at least in the large $N$ limit). 
This can be shown as follows. Let us rescale the field 
$A^a \rightarrow g A^a$. Then the action \eqn{act}
acquires the schematic form\footnote{Strictly speaking we must assume at 
this point that $g$ is small, and that $k g^2$ is of order 1. 
At the end of the day we will find that $g\sim 1/N$  and $k \sim N^2$, 
so our assumptions are indeed valid.}:
\brr 
S[A] &=&  \int_{B_5} \frac{1}{2} \d_{ab} \,
dA^a\wedge {}^*dA^b  + g S[A]_{\rm 3-pt}\ .
\label{act-sch}
\err 
We may write the field $A^a$ which solves the {\it full} equations of motion 
as
\bq A^a = A^{a(0)} + g A^{a(1)} + \Oo (g^2) \ .
\label{ansatz}
\eq 
By $A^{a(0)}$ we mean the field we introduced above in \eqn{valueA}. It
satisfies the free 
equations of motion, and approaches the desired boundary field 
$\Aa^a$ when $x_0$ goes to zero. This implies that we must put 
$A^{a(1)}$ equal to zero at the boundary of $\mbox{AdS}_5$. Given the ansatz
\eqn{ansatz} the action \eqn{act-sch} becomes
\brr 
S[A] &=& \int_{B_5} \frac{1}{2} \d_{ab} \,
dA^{a(0)}\wedge {}^*dA^{b(0)} + g \,\d_{ab} \,
dA^{a(1)}\wedge {}^*dA^{b(0)} + g \, S[A^{(0)}]_{\rm 3-pt}\ ,
\label{act-sch2}
\err 
where we dropped terms of order $g^2$. 
The $A^{(1)}$ dependent term, however, does not contribute, as can be seen 
by performing a partial integration which removes the derivative from 
$A^{(1)}$. This generates a bulk term which vanishes due to the fact that
$A^{(0)}$ satisfies the free equation
of motion. In addition there is a boundary term which vanishes due to the 
boundary condition on $A^{(1)}$.   

In summary, we see that it suffices to insert the relations \eqn{valueA}
into the action \eqn{act}. As such we can write down explicit integral 
forms of the supergravity action up to cubic powers of the boundary 
value $\Aa_i$.  From this
we will then determine the two- and three-point functions of $\cN=4$ $SYM$
$R$-currents, by taking the appropriate functional derivatives with respect to
the boundary values $\Aa_i$.

\subsection{Two-point function.}

The quadratic action
$S_2[A]$ is the relevant part for the two-point
correlator,
\brr S_2 &=&  \frac{1}{2g^2} \int d^d\xb_1 d^d\xb_2 ~\Aa^a_i(\xb_1)
     \Aa^b_j(\xb_2)~ \delta_{ab} \nonu
     && \times\hspace{2mm} \bl (\der_1^i \der_1^j - \d^{ij} \der_1^2)
     I^{d/2-1}_{d-2,\,d-2}
     - {1\over 8(d-2)^2} (\der_1^i \der_1^j - \delta^{ij} \der_1^2 )
     \der_1^2  I^{d/2-2}_{d-3,\,d-3} \br\ .
\label{twopoint}
\err
Here we have used integrals $I^f_{mn}$ defined as
\brr I^f_{mn} (\xb_1,\xb_2) = \int dx_0 \, d^d\xb
     \frac{x_0^{2f+1}}{R(\xb_1;x_0,\xb)^{m+1} R(\xb_2;x_0,\xb)^{n+1}} \, .
\label{Iinttwo}
\err
These integrals converge only for $n+m>f+(d/2-1)$, which is satisfied in the
case at hand. We have simplified the expression \eqn{twopoint} somewhat
by using the fact that the two-point function is
translationally invariant on the boundary. This implies that the integrals
$I^f_{mn}(\xb_1,\xb_2)$ are functions of the difference $\xb_{12}\equiv
\xb_1-\xb_2$ only, and that $\der_2$ acting in \eqn{twopoint} may be
converted into $- \der_1$.

The evaluation of the correlator involves the computation
of the integrals $I^f_{mn}(\xb_1,\xb_2)$ in eqs. \eqn{Iinttwo}.
In the  following we take the dimension to be $d=4$ because the result 
is finite for separated points.  
After Feynman parameterization and performing the
$x_0,\bar{x}$  integration we obtain
\brr 
I^f_{mn}(\xb_1,\xb_2)
      = {\Lam^f_{mn}}\frac{1}{\xb_{12}^{2(m+n-f-1)}} \int_0^1 d\a_1~
      {\a_1^{f+1-n}(1-\a_1)^{f+1-m}}
\err
where
\bq \Lam^f_{mn}={\Omega(4)\over 4}
      {\G(f+1) \G(m+n-f-1)\over \Gamma(m+1)\Gamma(n+1)}
\eq
and the constant $\Omega(4)=2\pi^2$ is the angular integral over
four-dimensional space.

Notice that in the second term of~\eqn{twopoint} the derivative 
$\der_1^2 I^0_{11}(\xb_1,\xb_2)$ produces
\brr
\der_1^2
\frac{1}{\xb_{12}^{2}} = -4\pi\d^4(\xb_1-\xb_2) \ ,
\err
which is just the four-dimensional propagator.  We
are interested in the case of separated points $\xb_1 \neq \xb_2$,
and we will thus ignore the above contact contribution.

In this way one finds for the two-point function
\brr
\langle J^i_a(\xb_1) J^j_b(\xb_2) \rangle &\equiv& -
      \frac{\d}{\d\Aa^a_i (\xb_1) } \frac{\d}{\d\Aa^b_j
      (\xb_2) } S_{2}[A(\Aa)]  \Big|_{\Aa=0} \nonu
      &=& -\frac{1}{g^2} \delta_{ab}
       (\der_1^i \der_1^j - \d^{ij} \der_1^2) I^{1}_{22}
     \nonu
      &=& \frac{3 \pi^2}{2g^2} \d_{ab} {1\over {\bar x}_{12}^6}
      I^{ij}(\xb_{12}) \ .
\label{prop}
\err
Here and in what follows we use the shorthand
\bq I^{ij}(\xb)=  \delta^{ij} - 2 {\xb^i \xb^j\over \xb^2}  \ .
\eq
Equation \eqn{prop} is the well-known unique conformal invariant
two-point function for vector fields.

\subsection{Three-point function.}

Next we find the $\mbox{AdS}$ three-point correlator of
$SU(4)$ currents. The integrals are more complicated then for
the two-point function. We will not calculate them explicitly, 
but only compute the leading term in the short distance
expansion (OPE) and then determine the full correlator using
conformal invariance.  A proof that the $\mbox{AdS}$ theory leads 
to conformally covariant correlators has been given in \cite{kall}.
The full three-point function is defined via
\bq \langle J^i_a(\xb_1) J^j_b(\xb_2) J^k_c (\xb_3) \rangle
    = -\frac{\d}{\d\Aa^a_i (\xb_1) } \frac{\d}{\d\Aa^b_j
    (\xb_2) } \frac{\d}{\d\Aa^c_k (\xb_3) } S_3[A(\Aa)]\Big|_{\Aa=0}
 \ .
\eq
There are two separate contributions,
distinguished  by the group theory factor $d_{abc}$ or
$f_{abc}$.   The Yang-Mills portion of the supergravity action
yields
\brr
S_{3, f_{abc}} &=& \frac{1}{2g^2}
\int d^d \xb_1 \, d^d \xb_2 \, d^d \xb_3\,
     \Aa_i^a (\xb_1)  \Aa_j^b (\xb_2) \Aa_k^c (\xb_3) f_{abc}\nonu
     &&\times\hspace{2mm}\bl 2 \d_{ij} \,\der_{1k} I^{d-2}_{d-2,d-2,d-2}
     + \frac{1}{2(d-2)^2} \der_{2j}  (\der_{1i} \der_{1k} -
     \d_{ik}\der_1^2 ) I^{d-3}_{d-3,d-3,d-2} \br .
\label{actkin}
\err
The expression in eq.~\eqn{actkin} can still be anti-symmetrized
in the indices in accord with the $f_{abc}$ structure.
The integrals $I^f_{mnp}(\xb_1,\xb_2,\xb_3)$
are defined
analogously to the $I^f_{mn}(\xb_1,\xb_2)$ we used for the two-point
function 
\bq I^f_{mnp} (\xb_1,\xb_2,\xb_3) = \int dx_0 \, d^d\xb
    \frac{x_0^{2f+1}}{R(\xb_1;x_0,\xb)^{m+1}
    R(\xb_2;x_0,\xb)^{n+1}R(\xb_3;x_0,\xb)^{p+1}} \ .
\label{Iintthree}
\eq
These integral expressions are convergent for $m+n+p > f +(d/2-2)$. All
the integral expressions involved in the correlator satisfy this condition.
We maintain separation of points.
The Yang-Mills AdS contribution to the correlator is (with $d=4$), 
\brr && \hspace{-.7cm} \langle J^i_a(\xb_1) J^j_b(\xb_2) J^k_c
(\xb_3)
     \rangle_{f_{abc}} \nonu
     &=& -\frac{f_{abc}} {g^2} \Bigl\{ [ \d_{ij} (\der_{1k} -
\der_{2k} )
     +\d_{jk} (\der_{2i} - \der_{3i} )+ \d_{ki} (\der_{3j} - \der_{1j} ) ]
     I^2_{222} \nonu
     && \phantom{\frac{f_{abc}} {g^2} \Bigl\{ }
     +\frac{1}{16} [ \der_{2j} (\der_{1i}\der_{1k} - \d_{ik}\der_1^2)
     - \der_{1i} (\der_{2j}\der_{2k} - \d_{jk}\der_2^2 ) ] I^1_{112}
     \nonu
     &&\phantom{\frac{f_{abc}} {g^2} \Bigl\{ }
     +\frac{1}{16} [ \der_{3k} (\der_{2i}\der_{2j} - \d_{ij}\der_2^2 )
     -\der_{2j} (\der_{3i}\der_{3k} - \d_{ik} \der_3^2)] I^1_{211}
     \nonu
     &&\phantom{\frac{f_{abc}} {g^2} \Bigl\{ }
     + \frac{1}{16} [ \der_{1i} (\der_{3j}\der_{3k} - \d_{jk}\der_3^2 )
     - \der_{3k} (\der_{1i}\der_{1j} - \d_{ij} \der_1^2 )]
     I^1_{121} \Bigr\} \ .
\label{kin}
\err
The Chern-Simons component of the supergravity action reads,
\brr S_{3, d_{abc}} &=& \frac{ik}{32\pi^2} \int d^4 \xb_1 \,
d^4 \xb_2 \,
     d^4 \xb_3 \,
     \Aa_i^a (\xb_1)  \Aa_j^b (\xb_2)
     \Aa_k^c (\xb_3) \,d_{abc}\, \ve^{jklm} \frac{1}{d-2} \nonu
     &&\times\hspace{2mm}  \bl -\frac{1}{2}
     \der_{1i} \der_{2m} \der_{3l} I^{3d/2 -4}_{d-3,d-2,d-2}
     +  \der_{3l} (\der_{1i} \der_{1m} - \d_{im} \der_1^2 )
     I^{3d/2-4}_{d-3,d-2,d-2}\br \ .
\label{actcs}
\err
with symmetrized contribution to the three-point
function 
\brr && \hspace{-.7cm} \langle J^i_a(\xb_1) J^j_b(\xb_2) J^k_c
 (\xb_3)
       \rangle_{d_{abc}}\nonu
      &=& -\frac{ik\, d_{abc}}{64 \pi^2}  \Big\{
\epsilon_{jklm}[(\der_{1i}\der_{1l}-\delta_{il}
      \der_{1}^{2}) (\der_{2m}-\der_{3m})+\der_{1i}\der_{2l}\der_{3m}]
      I_{122}^{2}\nonu
      &&\phantom{ \frac{N\, d_{abc}}{64 \pi^2}   \{  }
      +\epsilon_{kilm}[(\der_{2j}\der_{2l}-\delta_{jl}\der_{2}^{2})
      (\der_{3m}-\der_{1m})+ \der_{2j}\der_{3l}\der_{1m}]
      I_{212}^{2} \nonu
      && \phantom{ \frac{N\, d_{abc}}{64 \pi^2}   \{  }
      +\epsilon_{ijlm}[(\der_{3k}\der_{3l}-\delta_{kl}\der_{3}^{2})
      (\der_{1m}-\der_{2m})+ \der_{3k}\der_{1l}\der_{2m}]I_{221}^{2} \Big\}
\ .
\label{dcorr}
\err
We will now examine in more detail the integral expressions 
occuring within the correlators.   

The Feynman parameterized integrals
are

\bq I^f_{mnp} (\xb_1,\xb_2,\xb_3) =   \Lambda^f_{mnp}  \int^\infty_0 d\a_1
    d\a_2 d\a_3
    \frac{\a_1^m \a_2^n \a_3^p ~\d(\a_1+\a_2+\a_3-1) }{\left\{\a_1\a_2
    \xb_{12}^2 + \a_1\a_3\xb_{13}^2
    + \a_2 \a_3\xb_{23}^2 \right\}^{m+n+p-f}} \ .
\eq
The factor $\Lambda^f_{mnp}$ equals
\bq \Lambda^f_{mnp} = {\Omega(4)\over 4} \frac{\Gamma(f+1)
    \Gamma(m+n+p-f)}{\Gamma(m+1)\, \Gamma(n+1)\, \Gamma(p+1)} \ .
\eq
It is convenient to change the parametric variables to $u$ and $w$ through
\bq \a_1 = \frac{1}{2}(1-u)(1+w), \hspace{2cm} \a_2 = \frac{1}{2}(1-u)
(1-w) \ .
\eq
The integral $I_{mnp}^f(\xb_1,\xb_2,\xb_3)$ then reduces to
\brr I^f_{mnp} &=& 2^{p-f-1} \Lambda^f_{mnp}  \int^{1}_{-1} dw \,
     (1+w)^m (1-w)^n \, H^{p;f-p+1}_{m+n+p-f} \ .
\label{wint}
\err
where the integrals $H^{p;q}_{r}$ are given by
\brr H^{p;q}_{r} &=& \int^{1}_{0} du \,\frac{u^p (1-u)^q}{(uP+Q)^r}
\label{defH}
\err
and
\brr P &=& -\frac{1}{2} (1-w^2) \xb_{12}^2 + (1-w) \xb_{23}^2 +
     (1+w) \xb_{13}^2 ~=~ \frac{1}{2} ( \xb_{13} + \xb_{23} + w
     \xb_{12} )^2 \nonu
     Q &=& {1\over 2}(1-w^2) \xb_{12}^2 \ .
\err
The result with general $p$ and $q$ for the integrals $H^{p;q}_{r}$ is
unwieldy, so we specialize to those cases which actually occur
within the correlator expression: $H^{2;1}_{4}$,
$H^{2;0}_{3}$, $H^{1;1}_{3}$, $H^{2;1}_{3}$ and $H^{1;2}_{3}$.
The $H^{2;1}_{4}$ integral may be expressed as 
\bq
H^{2;1}_{4} = -
\frac{1}{P} H^{2;0}_{3} +
\frac{2}{3P} H^{1;0}_{3} \ .
\eq
Moreover, it is easy to see by expanding out the factor $(1-u)^q$ of eq.
\eqn{defH} that the required $H^{p;q}_{3}$ integrals are composed
of the following building blocks:
\brr 
H^{1;0}_{3} &=& -\frac{1}{2} \frac{1}{P(P+Q)^2} - \frac{1}{2}
     \frac{1}{P^2(P+Q)} +\frac{1}{2} \frac{1}{P^2Q} \nonu
     H^{2;0}_{3} &=& -\frac{1}{2} \frac{1}{P(P+Q)^2} - \frac{1}{P^2(P+Q)}
     - \frac{1}{P^3} \ln \frac{Q}{P+Q}  \nonu
     H^{3;0}_{3} &=& -\frac{1}{2} \frac{1}{P(P+Q)^2} - \frac{3}{2}
     \frac{1}{P^2(P+Q)}
     + \frac{3}{P^3} + 3 \frac{Q}{P^4} \ln \frac{Q}{P+Q} \ .
\label{build}
\err
Further evaluation of the $w$ integration in eq.\eqn{wint} is involved.
Notice that these integrals gives rise to both logarithms and
dilogarithms. Conformal invariance, however, prohibits such functional forms 
from entering into the final result for the correlator. This
implies that  various problematic contributions coming from the
different $I_{mnp}^f$ integrals in eqs. \eqn{kin} and \eqn{dcorr} must
cancel out.  We will confirm this cancellation. 

Our strategy to compute the OPE of the correlation function 
consists of first acting with the derivatives appearing in
\eqn{kin} and
\eqn{dcorr} on the integrals $H^{p;0}_{3}$ .
It is convenient to use the following identity 
\bq \der_i H^{p;0}_{3} = - (p+1) \frac{\der_i P}{P} H^{p;0}_{3}
    - p ~\frac{\der_i Q}{P} H^{p-1;0}_{3}
    + \frac{\der_i (P+Q)}{P(P+Q)^3}
    - \delta_{p,0} \frac{\der_i Q}{PQ^3} \ ,
\label{actder}
\eq
which is valid for all $p\geq 0$. The integral $H^{0;0}_{3}$
appears within the correlator after acting with $\der_1$
or $\der_2$ derivatives on the integrals $H^{p;0}_{3}$. Notice that
$\der_3$ doesn't lower the value of $p$, because $\der_3 Q=0$. In addition 
we have that 
\brr H^{0;0}_{3} &=& -\frac{1}{2} \frac{1}{P(P+Q)^2} +\frac{1}{2}
\frac{1}{PQ^2} \ .
\err

With these formulae it is straigthforward to 
extract the leading short-distance behavior (OPE) of the correlation function.
In the limit $\xb_{12}\rightarrow 0$ one finds that
$P \propto (P+Q) \propto \xb_{23}^2$, whereas $Q \propto \xb_{12}^2$.
Therefore
the most singular terms originate from $H^{0;0}_{3}$ or derivatives
thereof. The reader can verify that in intermediate steps of the computation
order $\Oo (\xb_{12})^{-4}$ terms appear, but these cancel out in the
final result. At the order $\Oo (\xb_{12})^{-3}$ one finds for the $f_{abc}$
part of the correlator
\brr
\mathop{\lim_{\xb_{12}\rightarrow 0}} \langle J^i_a(\xb_1) J^j_b(\xb_2) J^k_c
(\xb_3)
\rangle_{f_{abc}} &=& -\frac{\Omega(4)}{2^5 g^2}
{f^{abc} \over \xb_{12}^4 \xb_{23}^6} \left\{  \a_2\bigg( \xb_{12}^i
I^{jk}(\xb_{23})+\xb_{12}^j
I^{ik}(\xb_{23})\bigg) \right. \nonu
&+&\left. \xb_{12}^l\bigg( \a_1{\xb_{12}^i \xb_{12}^j\over
\xb_{12}^2}-\a_2\delta^{ij}
\bigg) I^{kl}(\xb_{23}) \right\} \ ,
\label{leading}
\err
with coefficients $\a_1$ and $\a_2$ equal to $4$ and $5$
respectively.
The short-distance behavior of the Chern-Simons contribution is
\brr
\mathop{\lim_{\xb_{12}\rightarrow 0}} \langle J^i_a(\xb_1) J^j_b(\xb_2) J^k_c
(\xb_3)
\rangle_{d_{abc}} &=& -\frac{3ik \, \Omega(4) }{ 2^7 \pi^2}
d_{abc} {1\over \xb_{12}^4 \xb_{23}^6}
\epsilon^{ij}_{~~pq}
\xb_{12}^q I^{pk} (\xb_{23}) \ .
\label{lead2}
\err
Using conformal invariance we will lift these short-distance expressions 
to the full correlator.

Reconstructing the three-point function is
straightforward.  
Following \cite{osborn1, osborn2} we parameterize the most general
three-point correlator for currents of conformal
weight three as
\brr \langle J^i_a(\xb_1) J^j_b(\xb_2) J^k_c (\xb_3) \rangle &=&
     {1\over \xb_{13}^6 \xb_{23}^6}
 I^i_{~m}(\xb_{13})I^j_{~n}(\xb_{23})
     t_{abc}^{mnk}(X_{12}) \ .
\label{deft}
\err
The argument $X_{12}$ is the special conformal invariant
combination
\bq X^k_{12} ={\xb_{13}^k \over \xb_{13}^2}-{\xb_{23}^k \over \xb_{23}^2} \ .
\eq
Conformal invariance requires $t_{abc}^{ijk}(X)$ to be a homogeneous
function satisfying
\brr t_{abc}^{ijk}(\lambda X) &=& \lambda^{-3}t_{abc}^{ijk}(X) \nonu
     D^i_{~m}D^j_{~n}D^k_{~\ell} t_{abc}^{mn\ell}(X) &=&
t_{abc}^{ijk}(D X) \ ,
\label{rel}
\err
for all elements $D^i_{~j}$ in $SO(4,2)$. In addition the
expression
\eqn{deft} must
be symmetric under the interchange of $(\xb_1, a, i)$ with $(\xb_2, b, j)$
or $(\xb_3, c, k)$ which puts further restrictions on the function
$t_{abc}^{ijk}(X)$. Solving for the most general tensor compatible with these
considerations leads to~\cite{osborn1, osborn2}
\bq t_{abc}^{ijk}(X) = f_{abc} \frac{1}{X^4}
    \left[ \beta_1 \frac{X^i X^j X^k}{X^2} + \beta_2 ( X^i \d^{jk} + X^j \d^{ik}
    -X^k \d^{ij}) \right] +  \beta_3 d_{abc} \epsilon^{ijkl}\frac{X^l}{X^4} \ ,
\label{tensor}
\eq
where $\beta_1$, $\beta_2$, $\beta_3$ are constant parameters which are model
dependent. For the leading short distance
behavior one finds that
\brr \langle J^i_a(\xb_1) J^j_b(\xb_2) J^k_c (\xb_3) \rangle &=&
     t_{abc}^{ijk'}(x_{12}) \frac{I^{k'k}(\xb_{23})}{\xb_{23}^6} +
     \mbox{subleading terms} \ ,
\label{formOPE}
\err
which means that the complete $t_{abc}^{ijk}(X)$ functions,
including the particular values for $\beta_1$, $\beta_2$ and $\beta_3$, can
be directly inferred from the OPE. From eqs.~\eqn{leading} and~\eqn{lead2} we find
\brr
\beta_1 = -\frac{4\Omega(4)}{2^5g^2} ,~&~ {\displaystyle \beta_2 =-
\frac{5\Omega(4)}{2^5g^2},}~&~
\beta_3 =-
\frac{3ik
\,
\Omega(4) }{ 2^7 \pi^2},
\err
which agrees with the result of \cite{freedman} up to an overall normalization.

As a check of the conformal invariance we also examined dilogarithmic
terms appearing in the expression for the correlator; the individual
integrals $I^f_{mnp}$  contain such terms. It is clear from eqs.
\eqn{deft} and \eqn{tensor} that the full correlator
is a rational function of the displacements $\xb_{12}$ and
$\xb_{23}$.  It is therefore essential that all the logarithmic and
dilogarithmic terms vanish.  We check the dilogarithms in the
$f_{abc}$ part only. Since $H^{3;0}_{3}$ doesn't appear in that
part of the correlator, it suffices to act with the derivatives of
eq. \eqn{kin} on $H^{2;0}_{3}$. One ends up with the following
problematic terms
\brr 
&&\hspace{-.7cm} \langle J^i_a(\xb_1) J^j_b(\xb_2) J^k_c (\xb_3)
      \rangle_{f_{abc}} \nonu
     &=&  \frac{3\pi^2 f_{abc}}{8 g^2} \int^{1}_{-1} dw \,
     (1-w^2)^2
     \frac{1}{P^5} \ln \frac{Q}{P+Q} M_{ijk} + \mbox{rational functions
     of}~ P,Q\nonu
\err
where
\brr 
M_{ijk} &=&(4+6 - 5(Y^2/P)) \bl 2w \d_{ij} Y_k + \d_{jk} Y_i
     (3-w) - \d_{ik} Y_j(3 + w) \br \nonu
     Y_i &=& \xb_{13i}  + \xb_{23i}+w \xb_{12i} \, .
\err
Given that $Y^2=2P$, the coefficient of the potential dilogarithmic terms
vanishes. 

\section{Momentum space}
\setcounter{equation}{0}
\label{mom}

We now proceed to compare the results in the supergravity calculation with those
from the super-Yang-Mills side of the correspondence.  As briefly discussed in
the introduction the regime where the supergravity calculation is valid is when
$g^2_{YM}N$ is large, contrary to the regime where perturbation theory  is valid
in super-Yang-Mills theory.  A priori it is therefore simplest to focus only on
the anomalous term of the three-point correlation function, where the
perturbative one-loop calculation must generate the complete answer.  The 
similar structure of the anomalous and the vector contributions to the 
correlator, however, leads to a simple one-loop correspondence for both 
terms.  The absence of higher-loop corrections to the correspondence 
indicates a non-renormalization theorem for the vector part of the 
correlator \cite{freedman}.  The comparison of these calculations will be made 
in momentum space; this requires us to Fourier transform the supergravity 
expressions.  Moreover the loop diagrams will in general be ultraviolet-divergent, 
and we use dimensional reduction to regularize our expressions.  

It is well known that closed expressions for loop integrals in
momentum space are complicated. We will therefore compare the
correlators at the integrand level. Because we do not have to
calculate the integrals explicitly this should be an effective way
to compare higher-point functions in the future. In particular the
supergravity side of the calculation becomes rather simple if left as an
integral expression.  

In the following we compute the correlation functions on the supergravity 
side in momentum space.  In the subsequent section we examine the 
correlators in the Yang-Mills theory and compare the two results 
with each other.  

\subsection{Two-point function: AdS}

We derive the two-point function from the Fourier transform of the
defining expression,
\brr
\langle J^i_a(\xb_1) J^j_b(\xb_2) \rangle &=&
       -\frac{\d_{ab}}{g^2} \left(
 (\der_1^i \der_1^j - \d^{ij} \der_1^2)
     I^{d/2-1}_{d-2,\,d-2}
     - {1\over 8(d-2)^2} (\der_1^i \der_1^j - \delta^{ij} \der_1^2 )
     \der_1^2  I^{d/2-2}_{d-3,\,d-3}  \right) \ , \nonu
\err
which leads to, 
\brr \langle J^i_a(q_1) J^j_b(q_2) \rangle
      &=&  \int d^4 \xb_1 d^4 \xb_2 e^{iq_1\xb_1+iq_2\xb_2}
      \langle J^i_a(\xb_1) J^j_b(\xb_2) \rangle \nonu
      &=& \frac{1}{g^2} \delta_{ab}
      (q_1^i q_1^j - \d^{ij} q_1^2)
      \bl I^{d/2-1}_{d-2,d-2} + \frac{1}{8(d-2)^2}\, q_1^2
I^{d/2-2}_{d-3,d-3} \br \ .
\label{twopmom}
\err
The partial derivatives have been integrated by parts and
have effectively been replaced by $-i$ times the
appropriate momentum.  The boundary terms vanish and one is left with
explicit functions of the momenta together
with the Fourier-transforms of the integral expressions $I^f_{mn}(\xb_1,\xb_2)$
given in eq. \eqn{Iinttwo}. These may easily be simplified to

\brr I^f_{mn}(q_1,q_2) &=& \int dx_0 d^d\xb d^d\xb_1 d^d\xb_2
       \frac{e^{iq_1\xb_1+iq_2\xb_2}x_0^{2f+1}}{(x_0^2 +|\xb-\xb_1|^2)^{m+1}
       (x_0^2 + |\xb-\xb_2|^2)^{n+1}}\nonu
      &=& \int dx_0 d^d\xb_1 d^d\xb_2~
      e^{iq_1\xb_1+iq_2\xb_2} \frac{x_0^{2f+1} \,(2\pi)^d
      \d^d(q_1+q_2)}{(x_0^2 + \xb_1^2)^{m+1}(x_0^2 + \xb_2^2)^{n+1}} \ .
\label{Ifmn}
\err
The momentum conserving delta function arises from
first shifting the $\xb_i$ integration variable to $\xb_i + \xb$
followed by an integration over $\xb$.  The boundary part of the
$\mbox{AdS}$-vertex position acts as a center-of-mass for the currents.

We shall evaluate the integrals through the use of multiple Schwinger
parameters.  Focusing on the $\xb_1$ integration we have

\brr 
\hskip -.3in 
I_m(x_0,q_1) = \int d^d \xb_1
      \frac{e^{iq_1\xb_1}}{(x_0^2+\xb_1^2)^{m+1}} 
      &=& {1\over\G(m+1)}  \int_0^{\infty} d\tau \tau^{m} \int d^d\xb_1
      e^{iq_1\xb_1-\tau(x_0^2+\xb_1^2)} \nonu
      &=& \frac{\Omega(d)}{2} \frac{\Gamma(d/2)}{\Gamma(m+1)}
      \int_0^{\infty} d\tau~\tau^{m-d/2} e^{-\tau x_0^2 - q_1^2/4\tau}\ ,
\label{ia}
\end{eqnarray}
where $\Omega(d)= 2 \pi^{d/2}/\Gamma(d/2)$ is the solid
angle in $d$ dimensions. One can perform the remaining
parameter integral to obtain 
\bq
I_m(x_0,q_1)=
\frac{2\pi^{d/2}}{\G(m+1)}\left(\frac{|q_1|}{2x_0}\right)^{m+1-d/2}
K_{m+1-d/2}(x_0|q_1|) \ ,
\eq
where $K_{\nu}(z)$ is the hyperbolic Bessel function of the second kind,
satisfying
\bq z^2\frac{\der^2}{\der^2 z}K_{\nu}(z)+z\frac{\der}{\der z}
      K_{\nu}(z)-(z^2+\nu^2)~ K_{\nu}(z)=0 \ .
\eq
The result, however, is not very useful as the final correlator
expression will then be a convolution of Bessel functions
 and such an integral is not easily done. We will instead keep
the integral form as in~\eqn{ia}.

The two-point integral function $I_{mn}^f(-q_1,-q_2)$ is obtained by
twice  substituting the result \eqn{ia} into eq. \eqn{Ifmn}, and we
arrive at, ignoring the overall $(2\pi)^d \delta^d(q_1+q_2)$,

\begin{eqnarray}
I^f_{mn}(q_1,q_2)&=&
{\pi^d\over\Gamma(m+1)\Gamma(n+1)} \int dx_0
d\tau_1d\tau_2 \, ~x_0^{2f+1} ~\tau_1^{m-d/2} \tau_2^{n-d/2}
e^{-x_0^2\tb - \sum q_r^2/4\tau_r} \nonu
&=&
{\pi^d\Gamma(f+1)\over2\Gamma(m+1)\Gamma(n+1)} \int
d\tau_1d\tau_2~ \frac{\tau_1^{m-d/2} \tau_2^{n-d/2}}{\tb^{f+1}}
e^{- \sum q_r^2/4\tau_r} \ ,
\label{I2}
\end{eqnarray}
where we have defined, 
\bq \tb = \sum_r \tau_r \ .
\eq

Rescaling $\tau_i \rightarrow \tau_i/4$ and substituting these integral
functions into the two-point correlator expression \eqn{twopmom} 
yields the dimensionally regularized  result
\brr &&
\hspace{-.7cm}\langle J^i_a(q_1) J^j_b(q_2) \rangle \nonu
      &=& \frac{ 8\,\pi^{2d}\, \G(d/2)}{g^2\, \G(d-1)^2} \,
      \delta_{ab} \,  \d^d(q_1+q_2)
      \nonu
      && \times \int d\tau_1 d\tau_2 ~\frac{(\tau_1
      \tau_2)^{d/2-2}}{\tb^{d/2}} e^{- \sum q_r^2/\tau_r}
      \Big[ q_1^i q_1^j - \d^{ij} q_1^2 \Big] \bpl 1+ q_1^2
      \frac{\tb}{(d-2)\tau_1\tau_2} \bpr \ .
\label{twoadsnf}
\err
We may rewrite the latter $q_1^2$ term as a derivative of the
exponent  and integrate by parts.  This manipulation will also be used in the 
derivation of the three-point function; in general this leads to  
\brr  \int_0^{\infty}d\tau_1
       \frac{\tau_1^{\a}}{\tb^{\beta}} \,q_1^2 e^{- \sum q_r^2/\tau_r}
      &=& \int_0^{\infty}d\tau_1
       \frac{\tau_1^{\a}}{\tb^{\beta}} \,\Big[ \beta\frac{\tau_1^2}{\tb}
- (\a+2) \tau_1 \Big] e^{- \sum q_r^2/\tau_r}
        \, .
\label{ident}
\err
Applying the identity \eqn{ident} within the integral \eqn{twoadsnf} we 
arrive at, 
\brr 
\langle J^i_a(q_1) J^j_b(q_2) \rangle
      &=& \frac{4\pi^{2d}}{g^2} \frac{\G(d/2)}{\G(d-1)^2} 
      \, \delta_{ab} \, \d^d(q_1+q_2) \nonu
      && \times \int d\tau_1 d\tau_2 ~\frac{(\tau_1\tau_2)^{d/2-2}}{\tb^{d/2}}
      e^{- \sum q_r^2/\tau_r}
      \Big[ q_1^i q_1^j - \d^{ij} q_1^2 \Big]  \ .
\label{twoads}
\err  
Since $q_1^2=q_2^2$ the exponent only depends on $1/\tau_1+1/\tau_2$.  After defining 
$\tau_1=\alpha\bar\tau$ and $\tau_2=(1-\alpha)\bar\tau$ we may scale the 
$\alpha$ parameters and perform the integration.  We obtain, 
\brr  
\langle J^i_a(q_1) J^j_b(q_2) \rangle 
= \frac{4\pi^{2d}}{g^2} \frac{\G(d/2)}{\G(d-1)^2} \, \delta_{ab} \,
\Big[ q_1^i q_1^j - \d^{ij} q_1^2 \Big] \int d\bar\tau \,\bar\tau^{d/2-3}
e^{-q^2/\bar\tau} \ ,  
\label{finaltwoads}
\err 
as the final integral expression for the two-point function.

\subsection{Three-point function: AdS}

Similar to the derivation of the momentum space two-point function, we
write the three-point correlator as polynomials in the momenta $q_i$ multiplied 
by the Fourier transformed integrals $I^f_{mnp}$.  The $f_{abc}$ part
of the correlator in eq. \eqn{kin} leads to  
\brr &&\hspace{-.7cm} \langle J^i_a(q_1) J^j_b(q_2) J^k_c (q_3)
      \rangle_{f_{abc}} \nonu
      &=& -\frac{i}{g^2}f_{abc} \ \bl \frac{1}{4(d-2)^2}
      \Big[ q_{2j} (q_{1i}q_{1k} - \d_{ik}q_1^2) - q_{1i} (q_{2j}q_{2k}
      - \d_{jk}q_2^2 ) \Big] I^{d-3}_{d-3,d-3,d-2} \nonu
      && \phantom{\frac{i}{g^2}f_{abc} \ \bl  }
      + \d_{ij} (q_{2k} - q_{1k} ) I^{d-2}_{d-2,d-2,d-2} + \mbox{cyclic} \br \ . 
\label{this}
\err
The momentum-space form of the expressions $I^f_{mnp}$ can be derived
using the methods of the previous section. Ignoring the
overall $(2\pi)^d \delta^d(q_1+q_2+q_3)$, we find 
\bq I^f_{mnp} = {\pi^{3d/2} \G(f+1)\over2\G(m+1)\G(n+1)\G(p+1)}
      \int d\tau_1d\tau_2d\tau_3 \,  \frac{\tau_1^{m-d/2}
      \tau_2^{n-d/2} \tau_3^{p-d/2}}{\tb^{f+1}} e^{-\sum q_r^2/4\tau_r} \ .
\label{If}
\eq
Inserting this result for $I^f_{mnp}$ into eq. \eqn{this} gives 
\brr 
&& \hspace{-.7cm} \langle J^i_a(q_1) J^j_b(q_2) J^k_c (q_3)
     \rangle_{f_{abc}} \nonu
     &=& -\frac{i8\, \pi^{5d/2}\G(d-2)}{g^2\,\G(d-1)^3} \ f_{abc} \
\delta^d(q_1+q_2+q_3)
     \int d\tau_1d\tau_2d\tau_3 \,
     \frac{(\tau_1\tau_2\tau_3)^{d/2-3}}{\tb^{d-1}} e^{-\sum q_r^2/\tau_r}
\nonu
     && \times \ \bl \tau_3 \tb \, \Big[ q_{2j} (q_{1i}q_{1k} - \d_{ik}q_1^2)
     - q_{1i} (q_{2j}q_{2k} - \d_{jk}q_2^2 ) \Big]  \nonu
     && \phantom{ \times\ \bl }
     + (d-2) \tau_1\tau_2\tau_3 \, \d_{ij} (q_{2k} - q_{1k} )
+\mbox{cyclic} \br \ .
\label{andthis}
\err
Again we write the
$q^2$ factors appearing within the tensor of eq. \eqn{andthis} as
$\tau$-derivatives of the exponent, followed by an integration by parts.
Using eq. \eqn{ident} in eq. \eqn{andthis}, and also taking into
account that the whole expression is invariant under cyclic permutations,
we arrive at
\brr && \hspace{-.7cm} \langle J^i_a(q_1) J^j_b(q_2) J^k_c (q_3)
     \rangle_{f_{abc}} \nonu
     &=& -\frac{i8\, \pi^{5d/2}\G(d-2)}{g^2\, \G(d-1)^3} \ f_{abc} \
     \delta^d(q_1+q_2+q_3)
     \int d\tau_1d\tau_2d\tau_3 \,
     \frac{(\tau_1\tau_2\tau_3)^{d/2-3}}{\tb^{d-1}} e^{-\sum q_r^2/\tau_r}
     \nonu
     && \times \ \Big[  \tau_3 \tb \, \Big( q_{1i} q_{2j} q_{1k} - q_{1i}
     q_{2j}q_{2k}
      \Big)  \nonu
     && \phantom{ \times\ \bl }
     + (d-2) \tau_1\tau_2 \, \d_{ij} \, \bpl \tau_3 (q_{2k} - q_{1k} )
     + (\tau_1 - \tau_2) q_{3k} \bpr
     +\mbox{cyclic} \Big] \ .
\label{finalads}
\err
We will compare this expression in section 4 to the corresponding 
one-loop calculataion in super-Yang-Mills theory.

Next we turn our attention to the $d_{abc}$ part of the three-point correlator. 
The integral expression in eq. \eqn{dcorr} leads to a momentum space form, 
\brr && \hspace{-.7cm} \langle J^i_a(q_1) J^j_b(q_2) J^k_c (q_3)
       \rangle_{d_{abc}}\nonu
      &=& \frac{k}{32 \pi^2 (d-2) } \, d_{abc} \,\bl \epsilon_{jklm}
      \bpl q_{1i}q_{1l}-\delta_{il}
      q_{1}^{2}\bpr (q_{2m}-q_{3m}) I_{d-3,d-2,d-2}^{3d/2-4} \nonu
      && \phantom{ \frac{iN}{32 \pi^2 (d-2) }\,  d_{abc} \,\bl  }
      + \epsilon_{jklm} \ q_{1i}q_{2l}q_{3m} \ I_{d-3,d-2,d-2}^{3d/2-4}
     + \mbox{cyclic}  \br \ .
\label{dcorrmom}
\err
After inserting the form \eqn{If} for $I_{mnp}^f$ into 
eq. \eqn{dcorrmom} we find the result 
\brr 
&& \hspace{-.7cm} \langle J^i_a(q_1) J^j_b(q_2) J^k_c (q_3)
     \rangle_{d_{abc}}\nonu
     &=& \frac{k\,  2^d \pi^{5d/2 -2} \, \G (3d/2 -3) }{16 \ \G
     (d-1)^3 }
     \, d_{abc}
     \ \delta^d(q_1+q_2+q_3 )  \int d\tau_1d\tau_2 d\tau_3 \,
     \frac{(\tau_1\tau_2\tau_3)^{d/2-2}}{\tb^{3d/2-3}} e^{-\sum q_r^2/\tau_r}
     \nonu
     &&\times \ \bl \epsilon_{jklm} \, \frac{1}{\tau_1} \,
     \Big[ \bpl q_{1i}q_{1l}-\delta_{il}
     q_{1}^{2}\bpr (q_{2m}-q_{3m})
     + \ q_{1i}q_{2l}q_{3m}   \Big]   + \mbox{cyclic}  \br \nonu
     &=& \frac{k}{4} \, \pi^{8}\, d_{abc}
     \ \delta^4(q_1+q_2+q_3 )  \int d\tau_1d\tau_2 d\tau_3 \,
     \frac{1}{\tb^{3}}\, e^{-\sum q_r^2/\tau_r}
     \nonu
     &&\times \ \bl \epsilon_{jklm} \, \frac{1}{\tau_1} \, q_{1i}
     \bpl  q_{1l} q_{2m} + q_{2l}q_{3m} + q_{3l} q_{1m} \bpr
     -\epsilon_{ijkm} \, \frac{1}{\tau_1} \,
     q_{1}^{2} (q_{2m}-q_{3m})  + \mbox{cyclic}  \br \ .
\label{further}
\err
Notice that in the last step we have set $d=4$ because the integral is
finite, as we expect for the anomaly. 

\section{Yang-Mills}
\setcounter{equation}{0}

Now that we have obtained expressions for the three-point functions from
supergravity tree graphs, we will calculate the same expressions directly
in ${\cN}=4$ $SU(N)$ super-Yang-Mills theory. At the end we shall compare the
momentum dependent results.  The ${\cal N}=4$ Yang-Mills theory follows 
from the dimensional reduction of $\cN=1$ super-Yang-Mills theory in ten 
dimensions and is
\brr
{\cal S}&= { Tr \displaystyle\int d^4x}& \frac{1}{4} F_{ij}^2 - \frac{i}{2}
\bar{\psi}_a\Ds\psi^a - \frac{1}{2} D_i\vphi_{ab}D^i\vphi^{ab}
-\frac{i}{2}\bar{\psi}_a [\vphi^{ab},\psi_b ] \nonu
&&
+\frac{1}{4}[\vphi_{ab},\vphi_{cd}][\vphi^{ab},\vphi^{cd}] \ .
\err
Under $R$-symmetry the fermions
transform chirally in the fundamental of $SU(4)$, whereas the scalars
transform as the antisymmetric $\bf{6}$, 
\brr
\d\psi^a = \ve^A (T_A)^a_{~b}\frac{(1+\g_5)}{2}\psi^b & ,& \d \vphi_{ab} =
\ve^A(T_A)_{ab}^{~~cd}\vphi_{cd} \ .
\err
The global $SU(4)$
current derived by the Noether method is
\brr
J_a^\mu(x) = \frac{1}{2}\vphi(x) T_a^{\vphi}
(\stackrel{\leftrightarrow}{\partial^\mu} +
2A^\mu(x))\vphi(x)
 - {i\over 2}\bar\psi(x) T_a^\psi \gamma^\mu\frac{(1+\gamma_5)}{2}\psi(x) \ .
\err
For the real representation {\bf 6} the generators are
antisymmetric; the quadra\-tic Casimirs are $C_4\equiv\frac{1}{2}$,
$C_6=1$.

\subsection{Two-point function: SYM}

The two-current correlator receives at one-loop contributions from two 
graphs, one with internal scalars $\vphi$ and one with fermions
$\psi$.  We first evaluate the  scalar contribution, 

\bq
\langle J^i_a(q_1) J^j_b(q_2)\rangle_{\vphi} =-\frac{1}{2}
(2\pi)^d\d^d(q_1+q_2){Tr_{6}(T_aT_b)}\int \frac{d^dp}{(2\pi)^d}
\frac{(2p+q_1)^i(2p+q_1)^j}{p^2(p+q_1)^2} \ ,
\label{2ym}
\eq
where we have dimensionally regularized the integral and suppressed a 
group theory factor of $N^2-1$.  It is trivial
to  perform the integration over the loop momentum in this particular case. We  will,
however, use the two-point function as a didactic tool for higher-point  ones, where
the loop integrals are not so straightforward. 

Introducing $SYM$ Schwinger parameters we may rewrite~\eqn{2ym} as, 
ignoring the overall factor of $(2\pi)^d\d^d(q_1+q_2)$,
\bq
\langle J^i_a(q_1) J^j_b(q_2)\rangle_{\vphi} =
\frac{C_6 \d_{ab}}{2}\int \frac{d^dp}{(2\pi)^d} \int_0^{\infty} d\tau_1d\tau_2~
(2p+q_1)^i(2p+q_1)^j e^{-\tau_1p^2-\tau_2(p+q_1)^2} \ .
\label{3ym}
\eq
The vertex factors $(2p+q_1)^i$ can be written as derivatives of the
exponential, after which the loop momentum integral is easily done:
\brr 
&& \hspace{-.7cm} \langle J^i_a(q_1) J^j_b(q_2)\rangle_{\vphi} \nonu
&=& \frac{C_6\d_{ab}}{2}
\int \frac{d^dp}{(2\pi)^d} \int_0^{\infty} d\tau_1d\tau_2
\left[
\left(\frac{1}{\tau_2}\frac{\der}{\der q_{1j}} + q_1^j\right)
\left(\frac{1}{\tau_2}\frac{\der}{\der q_{1i}} + q_1^i\right)
+\frac{\d^{ij}}{\tau_2}\right]
e^{-\tau_1p^2-\tau_2(p+q_1)^2} \nonu
&=&\frac{\pi^{d/2}C_6\d_{ab}}{2(2\pi)^d}\int_0^{\infty}
\frac{d\tau_1d\tau_2} {\bar{\tau}^{d/2}}~\left[
\left(\frac{1}{\tau_2}\frac{\der}{\der q_{1j}} + q_1^j\right)
\left(\frac{1}{\tau_2}\frac{\der}{\der q_{1i}} + q_1^i\right)
+\frac{\d^{ij}}{\tau_2}
\right]e^{-\frac{\tau_1\tau_2}{\bar{\tau}}q_1^2} \ ,
\label{4ym}
\err
where $\bar{\tau}=\tau_1+\tau_2$.  After expanding the derivatives we
arrive at
\bq
\langle J^i_a(q_1) J^j_b(q_2)\rangle_{\vphi} =
\frac{\pi^{d/2}C_6\d_{ab}}{2(2\pi)^d}\int_0^{\infty}
\frac{d\tau_1d\tau_2}{
\tb^{d/2}}~\left[
 \left({\tau_1-\tau_2\over\bt}\right)^2 q_1^iq_1^j + {2\over
\bt}\delta^{ij}  \right] e^{-{\tau_1\tau_2\over\bt}q_1^2} \ ,
\label{scatwo}
\eq
which is the final expression we need for the scalar contribution.

Next we turn to the fermionic contribution to the current-current
correlator; its integral expression is
\bq
\langle J^i_a(q_1) J^j_b(q_2)\rangle_{\psi}
=\frac{Tr_4(T_aT_b)}{4}
 {\rm Tr}\left[\gamma^i{(1+\gamma_5)}\gamma^k \gamma^j
{(1+\gamma_5)}
\gamma^{\ell} \right]\int {\frac{d^dp}{(2\pi)^d}}~  
 {p_k (p+q_1)_{\ell}\over p^2(p+q_1)^2} \ .
\label{fermpart}
\eq
We evaluate the momentum integral of eq.\eqn{fermpart} in a similar
way and use four dimensional commutation rules for the Dirac matrices
in accordance with the rules of dimensional reduction. This yields 
for the fermionic contribution, 

\brr
&& \hspace{-.7cm}
\langle J^i_a(q_1) J^j_b(q_2)\rangle_{\psi} \nonu
&=&-\frac{\pi^{d/2}C_4\d_{ab}}{2(2\pi)^d}
 {\rm Tr}\left[\gamma^i\gamma_k \gamma^j
{(1+\gamma_5)}
\gamma_{\ell} \right]\int_0^{\infty}\frac{d\tau_1d\tau_2}{
\bar{\tau}^{d/2}}~\left[\frac{1}{2
\bar{\tau}}\d^{k\ell}-\frac{\tau_1\tau_2}{\bar{\tau}^2}q_1^kq_1^{\ell}
\right] e^{-{\tau_1
\tau_2\over\bt}q_1^2} \nonu
&=&-\frac{2\pi^{d/2}C_4\d_{ab}}{(2\pi)^d} \int_0^{\infty}\frac{d\tau_1d
\tau_2}{\bar{\tau}^{d/2}}~\left[
\left(\frac{2-d}{2\bar{\tau}}+\frac{\tau_1\tau_2}{\bar{\tau}^2}q_1^2
\right)\d^{ij}
-2\frac{\tau_1\tau_2}{\bar{\tau}^2}q_1^iq_1^{j}\right] e^{-{\tau_1
\tau_2\over\bt}q_1^2} \ .
\label{fertwo}
\err
One would  like to see that the correlation functions~\eqn{scatwo}
and~\eqn{fertwo} are transverse, as one would expect for conserved currents.
Indeed the transversality will become manifest in the course of
bringing the $SYM$ expressions into a form that can be easily
compared  to the $\mbox{AdS}$ result in eq. \eqn{finaltwoads}.  

In a similar fashion to the steps performed in deriving the two-point 
function on the supergravity  side, we scale $\tau_1=\alpha\bar\tau$ and
$\tau_2=(1-\alpha)\bar\tau$.  The 
$\alpha$ integration within the scalar and fermionic contributions may be 
performed in order to compare our results, and  we find after changing 
variables $\bar\tau\rightarrow 1/\bar\tau$,  
\brr  
\langle J^i_a(q_1) J^j_b(q_2)\rangle_{\vphi} &=& 
\frac{\pi^{d/2}C_6\d_{ab}}{2(2\pi)^d}\int_0^{\infty}
d\bar\tau \, \tb^{d/2-3}~ \nonu && 
\hskip -.5in
\times \Bigl\{ \left[ B({d\over 2}-1,{d\over 2}-1) - 4 B({d\over 2},{d\over 2})
 \right] q_1^iq_1^j  + 2 B({d\over 2},{d\over 2}) \bar\tau \delta^{ij}
   e^{-q_1^2/\bar\tau}  
\Bigr\} \ , 
\err   
where $B(x,y)$ is the Beta function.  
Using the identity in eq. \eqn{ident} we may rewrite the latter term 
in the above and make the result manifestly transverse, 
\brr  
\langle J^i_a(q_1) J^j_b(q_2)\rangle_{\vphi} = 
\frac{\pi^{d/2}C_6\d_{ab}}{2(2\pi)^d} 
{4\G({d\over 2})^2\over (d-2)\G(d)} \left[q_1^iq_1^j-q_1^2 \delta^{ij}\right]  \,
\int d\bar\tau \, \bar\tau^{d/2-3} e^{-q_1^2/\bar\tau} \ .  
\label{scalars}
\err 
The fermionic contribution similarly gives 
\brr 
\langle J^i_a(q_1) J^j_b(q_2)\rangle_{\psi} = \frac{2\pi^{d/2}C_4\d_{ab}}{(2\pi)^d} 
\, {\G({d\over 2})^2\over \G(d)} \left[q_1^iq_1^j-q_1^2 \delta^{ij}\right]  \,
\int d\bar\tau \, \bar\tau^{d/2-3} e^{-q_1^2/\bar\tau} \ . 
\label{fermions}
\err  
The sum of eqs. \eqn{scalars} and \eqn{fermions} gives the two-point 
result on the super-Yang-Mills theory side. 

We now compare the two two-point functions to eachother, but first we must 
restore the group factor $N^2$ and coupling $\lambda^2$.  
Comparing the coefficients of the sum above with the $\mbox{AdS}$ result~\eqn{twoads} 
we find agreement if
\brr
\frac{4 \pi^{2d} \lambda^2 \G(d/2)}{g^2 \G(d-1)^2} &=& 
 2\pi^{d/2} {N^2} \frac{\G(d/2)^2}{\G(d)}    
 \left(\frac{C_6}{d-2}+2C_4\right) \nonu &=& 
 2\pi^{d/2} {N^2} \frac{\G(d/2)^2}{\G(d)} \, \left(\frac{d-1}{d-2}\right) \ . 
\label{normtwo}
\err
We will use this relation to fix the overall normalization of the correlation
functions.

\subsection{Three-point function: SYM}

In this section we analyze the one-loop three-point function; it arises from
a triangle graph with internal scalars or fermions.  The scalar 
contribution gives, 
\brr 
&& \hspace{-.7cm} \langle J_a^i(q_1) J_b^j(q_2)
     J_c^k(q_3)\rangle_{\phi}\nonu
     &=& i\frac{C_6 f_{abc}}{2} \d^{d} (q_1 + q_2 + q_3)
     \int d^d p \frac{ (2p-q_2+q_3)_i (2p-q_2)_j
     (2p+q_3)_k }{(p-q_2)^2 \, p^2 (p+q_3)^2} \ .
\err
The fermionic graph can be decomposed into an even and odd parity part:
\brr &&\hspace{-1.5cm} \langle J_a^i(q_1) J_b^j(q_2)
     J_c^k(q_3)\rangle_{\psi} \nonu
     && \hspace{-1.0cm}= - 2i C_4 \d^{d} (q_1+q_2+q_3)  \int d^d p \frac{(p-q_2)_l
     p_m (p+q_3)_n }{(p-q_2)^2 p^2 (p+q_3)^2}
     ( f_{abc} E_{iljmkn} - i d_{abc}O_{iljmkn}) \ ,
\label{fermm}
\err
with the tensors defined by 
\bq E_{iljmkn} =  {\rm Tr} \left\{ \g_i \g_l \g_j \g_m \g_k \g_n \right\} \ ,
    \hspace{2cm}  
    O_{iljmkn} =  {\rm Tr} \left\{ \g_i \g_l \g_j \g_m \g_k \g_n \g_5 \right\} \ .
\eq
As before the denominator factors may be exponentiated by means of a set of
Schwinger parameters, and the numerator factors can be generated by applying
derivatives on the exponentials.  The scalar graph then has the form
\brr &&\hspace{-.7cm}\langle J_a^i(q_1) J_b^j(q_2) J_c^k(q_3)
     \rangle_{\phi} \nonu
     &=& i\frac{C_6 f_{abc}}{2} \,\d^{d} (q_1+q_2+q_3)
     \int d^d p \int_0^{\infty} d\tau_1d\tau_2d\tau_3\, \nonu
     && \times \left[ - \left( \frac{1}{ \tau_3} \frac{\der}{\der q_2^j}
     + q_{2j} \right) \left( \frac{1}{ \tau_2} \frac{\der}{\der q_3^k}
     + q_{3k} \right) \left( \frac{1}{2\tau_3} \frac{\der}{\der q_2^i}
     -\frac{1}{2\tau_2}\frac{\der}{\der q_3^i}\right)
     - \d_{ij} \frac{1}{ \tau_3} \left( \frac{1}{ \tau_2}
     \frac{\der}{\der q_3^k} + q_{3k} \right)\right. \nonu
     && \phantom{\times [ }\ \left.  + \d_{ik} \frac{1}{ \tau_2}
     \left(\frac{1}{ \tau_3} \frac{\der}{\der q_2^j} + q_{2j} \right)
     \right] e^{-\tau_1 p^2 - \tau_2 (p+q_3)^2 - \tau_3(p-q_2)^2 } \ .
\err
At this point one can easily perform the integral over the loop momentum $p$
\bq \int d^d p \ e^{ -\tau_1 p^2 - \tau_2 (p+q_3)^2 -
    \tau_3(p-q_2)^2}
    = \left(\frac{\pi}{\tb}\right)^{d/2}
    e^{ -\frac{\tau_1\tau_2\tau_3}{\tb} ( q_2^2/\tau_2 + q_3^2/\tau_3
    +(q_2+q_3)^2/\tau_1)} \ ,
\eq
where we have defined $\tb = \tau_1+\tau_2+\tau_3$.  Expanding the 
derivatives yields
\brr 
&&\hspace{-.7cm}\langle J_a^i(q_1) J_b^j(q_2) J_c^k(q_3)
     \rangle_{\phi} \nonu
     &=& i\frac{C_6 f_{abc}}{2} \,\d^{d} (q_1+q_2+q_3)
     \int_0^{\infty} d\tau_1d\tau_2d\tau_3\,
     \frac{\pi^{d/2}}{\tb^{d/2+1}} \ e^{ -\frac{\tau_1\tau_2\tau_3}{\tb}
     \sum \tau_r^{-1} q_r^2 } \nonu
     && \times \left[ \ - \frac{1}{6 \tb^2}
     \bpl 2 \tau_1 q_{2i} + (\tb - 2\tau_2) q_{1i} \bpr
     \bpl 2 \tau_2 q_{3j} + (\tb - 2\tau_3) q_{2j} \bpr
     \bpl 2 \tau_3 q_{1k} + (\tb - 2\tau_1) q_{3k} \bpr \right. \nonu
     && \phantom{\times [ }\ +\frac{1}{6 \tb^2}
     \bpl 2 \tau_1 q_{3i} + (\tb - 2\tau_3) q_{1i} \bpr
     \bpl 2 \tau_2 q_{1j} + (\tb - 2\tau_1) q_{2j} \bpr
     \bpl 2 \tau_3 q_{2k} + (\tb - 2\tau_2) q_{3k} \bpr \nonu
     && \phantom{\times [ }\ \left. +\frac{2}{\tb} \d_{ij} \bpl \tau_3 (q_{2k}
     - q_{1k}) + (\tau_1 - \tau_2) q_{3k} \bpr + \mbox{cyclic} \right] \ .
\label{beforescale}
\err
The expression has been written in a manifestly cyclic form under the simultaneous
interchange of $1\rightarrow 2\rightarrow 3$ and $i \rightarrow j
\rightarrow k$ through the use of the momentum conservation. In addition, 
one may check that the expression in eq. \eqn{beforescale} is antisymmetric under the
simultaneous interchange of $1\leftrightarrow 2$ and $i\leftrightarrow j$.

In order to make the exponential
in \eqn{beforescale} equal to the one found on the $\mbox{AdS}$-side,
we impose the relation
\bq \tau^{{AdS}}_r =  \left[ \tau_r   \left( \frac{\bt}{ \tau_1\tau_2\tau_3}
    \right) \right]^{SYM} \ ,
\eq
which may be inverted to
\bq \tau^{SYM}_r =  \left[ \tau_r   \left( \frac{\bt}{ \tau_1\tau_2\tau_3}
    \right) \right]^{{AdS}} \ .
\eq
The Jacobian associated with this change of variables is 
\bq \left[ d\tau_1d\tau_2 d\tau_3 \right]^{SYM} =  \left[ d\tau_1d\tau_2
    d\tau_3 \left( \frac{\bt}{ \tau_1\tau_2\tau_3} \right)^3 \right]^{{AdS}}  
 \ .
\eq
In terms of the new variables we obtain for the scalar contribution, 
\brr 
&&\hspace{-.7cm}\langle J_a^i(q_1) J_b^j(q_2) J_c^k(q_3)
     \rangle_{\phi} \nonu
     &=& i\frac{C_6 f_{abc}}{2} \, \pi^{d/2} \d^{d} (q_1+q_2+q_3)
     \int_0^{\infty} d\tau_1d\tau_2d\tau_3\,
     \frac{ (\tau_1\tau_2\tau_3)^{d/2-3} }{\tb^d} \ e^{ - \sum \tau_r^{-1}
     q_r^2 } \nonu
     && \times \Big[ - \frac{1}{6} \bpl 2 \tau_1 q_{2i} + (\tb - 2\tau_2)
     q_{1i} \bpr
     \bpl 2 \tau_2 q_{3j} + (\tb - 2 \tau_3) q_{2j} \bpr
     \bpl 2\tau_3 q_{1k} + (\tb - 2\tau_1) q_{3k} \bpr \nonu
     && \phantom{\times [}\ + \frac{1}{6}
     \bpl 2 \tau_1 q_{3i} + (\tb - 2\tau_3) q_{1i} \bpr
     \bpl 2 \tau_2 q_{1j} + (\tb - 2\tau_1) q_{2j} \bpr
     \bpl 2 \tau_3 q_{2k} + (\tb - 2\tau_2) q_{3k} \bpr \nonu
     && \phantom{\times [}\ + 2 (\tau_1\tau_2\tau_3) \, \d_{ij}
     \bpl \tau_3 (q_{2k}-q_{1k}) + (\tau_1 - \tau_2) q_{3k} \bpr
     + \mbox{cyclic} \Big] \ .
\label{afterscale}
\err
Upon use of the identity, 
\bq \tau_1 \tau_3 \tb q_{1i} q_{2j} q_{1k}
     - \tau_2 \tau_3 \tb q_{1i} q_{2j} q_{2k} + \mbox{cyclic}
     = \tau_2 \tau_3 \tb q_{1i} q_{1j} q_{2k}
     - \tau_1 \tau_3 \tb q_{2i} q_{2j} q_{1k} + \mbox{cyclic} \ ,
\eq
we arrive at our final expression for the scalar graph
\brr &&\hspace{-.7cm}\langle J_a^i(q_1) J_b^j(q_2) J_c^k(q_3)
     \rangle_{\phi} \nonu
     &=& i\frac{C_6 f_{abc}}{2} \, \pi^{d/2} \d^{d} (q_1+q_2+q_3)
     \int_0^{\infty} d\tau_1d\tau_2d\tau_3\,
     \frac{ (\tau_1\tau_2\tau_3)^{d/2-3} }{\tb^d} \ e^{ - \sum \tau_r^{-1}
     q_r^2 } \nonu
     && \times \Big[ - 4 \tau_2 \tau_3^2 \bpl q_{1i} q_{2j} q_{1k}
     + q_{1i} q_{1j} q_{2k} \bpr
     + 4  \tau_1 \tau_3^2 \bpl q_{1i} q_{2j} q_{2k}
     + q_{2i} q_{2j} q_{1k} \bpr \nonu
     && \phantom{\times  [}\ - \frac{4}{3} \ \tau_1\tau_2\tau_3
     \bpl q_{2i} q_{3j} q_{1k} - q_{3i} q_{1j} q_{2k} \bpr
     + \tau_3 \tb^2 \bpl q_{1i}q_{2j}q_{1k} -q_{1i}q_{2j}q_{2k} \bpr \nonu
     && \phantom{\times [}\ + 2 (\tau_1\tau_2\tau_3) \, \d_{ij}
     \bpl \tau_3 (q_{2k}-q_{1k}) + (\tau_1 - \tau_2) q_{3k} \bpr
     + \mbox{cyclic} \Big] \ .
\label{fbos}
\err
Similarly, the even-parity
fermionic contribution can be written as
\brr &&\hspace{-.7cm}\langle J_a^i(q_1) J_b^j(q_2) J_c^k(q_3)
     \rangle_{\psi} {}_{\rm , even} \nonu
     &=&  i C_4 f_{abc} \pi^{d/2} \d^{d} (q_1+q_2+q_3)
     \int_0^{\infty} d\tau_1d\tau_2d\tau_3\,
     \frac{(\tau_1\tau_2\tau_3)^{d/2-3} }{\tb^d}
     e^{ - \sum \tau_r^{-1} q_r^2}  \ E_{iljmkn} \nonu
     &&  \times  \ \Big[ \ \frac{2}{3 } \left( \tau_1 q_{2l}
     -\tau_2 q_{1l}\right) \left( \tau_2 q_{3m}-\tau_3 q_{2m}\right)
     \left( \tau_3 q_{1n} -\tau_1 q_{3n}\right) \nonu
     && \phantom{ \times  \ [} \  - (\tau_1\tau_2\tau_3) \, \d_{lm}
     \left(\tau_1 q_{3n} - \tau_3 q_{1n} \right) +\mbox{cyclic} \Big]
\err
It is straightforward to work out the trace over
the gamma matrices using four dimensional rules and contract the
Lorentz indices.  As before we absorb any
$q_r^2$ prefactor into $exp\{- \sum \tau_r^{-1} q_r^2\}$; doing so, we find
\brr &&\hspace{-.7cm}\langle J_a^i(q_1) J_b^j(q_2) J_c^k(q_3)
     \rangle_{\psi} {}_{\rm , even} \nonu
     &=&  i C_4 f_{abc} \, \pi^{d/2} \d^{d} (q_1+q_2+q_3)
     \int_0^{\infty} d\tau_1d\tau_2d\tau_3\,
     \frac{(\tau_1\tau_2\tau_3)^{d/2-3} }{\tb^d}
     e^{ - \sum \tau_r^{-1} q_r^2 }\nonu
     && \times \  \Big[ 4 \tau_2 \tau_3^2 \bpl q_{1i} q_{2j} q_{1k}
     + q_{1i} q_{1j} q_{2k} \bpr
     - 4  \tau_1 \tau_3^2 \bpl q_{1i} q_{2j} q_{2k}
     + q_{2i} q_{2j} q_{1k} \bpr \nonu
     && \phantom{\times  [} + \frac{4}{3} \ \tau_1\tau_2\tau_3
     \bpl q_{2i} q_{3j} q_{1k} - q_{3i} q_{1j} q_{2k} \bpr\nonu
     && \phantom{\times  [}  + 2\tau_1\tau_2 \bl (\frac{d}{2}-1)\tb -
     \tau_3 \br
     \, \d_{ij} \bpl \tau_3 (q_{2k}-q_{1k}) + (\tau_1 - \tau_2)
     q_{3k} \bpr +\mbox{cyclic} \Big] \ .
\label{fferm}
\err

The bosonic and fermionic expressions
\eqn{fbos} and \eqn{fferm} benevolently conspire to produce a
simple  formula for the $f_{abc}$ part of the three-point
correlator; their sum is  
\brr &&\hspace{-.7cm}\langle J_a^i(q_1) J_b^j(q_2) J_c^k(q_3)
     \rangle_{f_{abc}} \nonu
     &=& i \frac{ f_{abc}}{2} \, \pi^{d/2} \d^{d} (q_1+q_2+q_3)
     \int_0^{\infty} d\tau_1d\tau_2d\tau_3\,
     \frac{(\tau_1\tau_2\tau_3)^{d/2-3} }{\tb^{d-1}}
     e^{ - \sum \tau_r^{-1} q_r^2 }\nonu
     && \times \  \Big[ \tau_3 \tb \bpl q_{1i}q_{2j}q_{1k}
    -q_{1i}q_{2j}q_{2k} \bpr \nonu
     && \phantom{\times  [}  + (d-2) \tau_1\tau_2 \,
     \d_{ij} \bpl \tau_3 (q_{2k}-q_{1k}) + (\tau_1 - \tau_2)
     q_{3k} \bpr +\mbox{cyclic} \Big] \ .
\label{total}
\err
Comparing with the result of the supergravity calculation \eqn{finalads} we see that
the two expressions agree. Restoring the suppressed group factor of $N^2-1$ and the 
boundary value to source normalization $\lambda$ the
overall normalizations are related as
\bq
i\frac{\pi^{d/2}}{2}N^2 = -i\frac{8\, \pi^{5d/2} \lam^3 \G(d-2)}{g^2\,\G(d-1)^3}
\ .
\label{normthree} 
\eq
In addition the normalizations of the two-point
function must agree according to eq.~\eqn{normtwo}. These two requirements fix the
undetermined normalizations
$\lam$ and
$g^2$ of the
$\mbox{AdS}$/CFT correspondence to 
\brr
\lam = -\frac{1}{8}\frac{(d-2)^2}{\pi^{d/2}}\frac{\G(d-1)}{\G(d/2)}~&,~ &~
g^2= \frac{\pi^{d/2} (d-2)^6}{32N^2}\frac{\G(d-2)}{\G(d/2)^3} \ .
\err
We should comment that the second equation does not imply that the value of the
coupling constant of the IIB supergravity has been fixed. As we explained in section 2,
$g^2$ is an additional normalization of the action. The expansion parameters of the
supergravity theory, $\a^{\prime}$ and $g_{st}$, are still arbitrary.

The odd parity part of the fermionic triangle graph is described by the same
formula~\eqn{fermm} as for the even one but with the gamma-matrix
trace, $O_{ijklmn}$.  The trace over the symmetrized set of gamma matrices
is calculated with four-dimensional rules,
\brr  O_{ijklmn}
     &=&\frac{1}{6} \mbox{Tr} \bl \g_i\g_l\g_j\g_m\g_k\g_n \g_5
     -\g_n\g_i\g_l\g_j\g_m\g_k \g_5+\g_k\g_n\g_i\g_l\g_j\g_m \g_5
     \nonu
     &&\phantom{ \frac{1}{6} \mbox{Tr} \bl} -\g_m\g_k\g_n\g_i\g_l\g_j \g_5
     +\g_j\g_m\g_k\g_n\g_i\g_l \g_5 -\g_l\g_j\g_m\g_k\g_n\g_i \g_5 \br
     \nonu
     &=&\frac{2}{3} ( -\d_{il}\e_{jkmn}+\d_{in}\e_{jklm}
     +\d_{jl}\e_{kimn} -\d_{jm}\e_{kinl}+\d_{km}\e_{ijnl}-\d_{kn}\e_{ijklm} )
     \nonu
     && - \frac{1}{3}
     (\d_{mn}\e_{ijkl}+\d_{lm}\e_{ijkn}+\d_{ln}\e_{ijkm})  \ .
\err
The evaluation of the triangle diagram leads to the integral
form,
\brr &&\hspace{-.7cm}\langle J_a^i(q_1) J_b^j(q_2) J_c^k(q_3)
     \rangle_{\psi} {}_{\rm , odd} \nonu
     &=&  -2 {C_4 d_{abc}} \, \pi^{d/2} \d^{d} (q_1+q_2+q_3)
     \int_0^{\infty} d\tau_1d\tau_2d\tau_3\,
     \frac{(\tau_1\tau_2\tau_3)^{d/2-2} }{\tb^d}
     e^{ - \sum \tau_r^{-1} q_r^2 }  O_{ijklmn}\nonu
     && \times \  \Big[\frac{1}{2} \d_{lm}A_n +
     B_{lmn} + \mbox{cyclic} \Big] \ ,
\label{dferm}
\err
where the tensors are defined as 
\brr A_n &=& \tau_1 q_{3n}-\tau_3 q_{1n} \nonu
     B_{lmn} &=& \tau_3 q_{1n}q_{2m}(\frac{q_{2l}}{\tau_2}-\frac{
     q_{1l}}{\tau_1})
     + \frac{1}{3} ( q_{1l}q_{2m}q_{3n}-q_{2l}q_{3m}q_{1n}) \ .
\err
The contraction of the $A$ term with the trace over
the gamma matrices is straightforward.
The $B$ terms contribute, in addition to an $\e_{ijkl}$ piece, the following
\brr
O_{ijklmn} B_{lmn} &=&
\frac{2}{3}\e_{jklm}[\tau_3 q_{1l}q_{2m}+\tau_1 q_{2l}q_{3m}+\tau_2 q_{3l}q_{
1m}][\frac{q_{3i}}{\tau_3}+\frac{q_{2i}}{\tau_2}-2\frac{q_{1i}}{\tau_1}]
+\mbox{cyclic}
\nonu
&&+ \epsilon_{ijkl} v^l \ .
\err
This contribution may be rewritten using Schoutens identity, $\e_{ijkl}q_m$ +
cyclic$\{ijklm\}$=0, into the form,
\brr
O_{ijklmn} B_{lmn} &=&
-2(\tau_3 q_{1l}q_{2m}+\tau_1 q_{2l} q_{3m}+\tau_2 q_{3l} q_{1m})(\e_{jklm}
\frac{q_{1i}}{\tau_1} +\mbox{cyclic}) \nonu
&&+ \epsilon_{ijkl} v^l \ .
\err
By appropriately extracting $q^2$ factors out of the exponent by means of
the identity \eqn{ident} we obtain the final result for the anomalous 
part of the triangle graph, 
\brr &&\hspace{-.7cm}\langle J_a^i(q_1) J_b^j(q_2) J_c^k(q_3)
     \rangle_{\psi} {}_{\rm , odd} \nonu
     &=&  4 {C_4 d_{abc}} \, \pi^{d/2} \d^{d} (q_1+q_2+q_3)
     \int_0^{\infty} d\tau_1d\tau_2d\tau_3\,
     \frac{(\tau_1\tau_2\tau_3)^{d/2-2} }{\tb^{d-1}}
     e^{ - \sum \tau_r^{-1} q_r^2 } \nonu
     && \times [ (\tau_3 q_{1l}q_{2m}+\tau_1 q_{2l} q_{3m}
     +\tau_2 q_{3l} q_{1m}) \e_{jklm}\frac{q_{1i}}{\tau_1 \tb}-\e_{ijkl}
     \frac{1}{(d-1) \tau_3} q_3^2 (q_1-q_2)_l
     +\mbox{cyclic}] \nonu
     &=&  \frac{4}{3} {C_4 d_{abc}} \, \pi^2 \d^4 (q_1+q_2+q_3)
     \int_0^{\infty} d\tau_1d\tau_2d\tau_3\,
     \frac{1}{\tb^3}
     e^{ - \sum \tau_r^{-1} q_r^2 } \nonu
     && \times [ ( q_{1l}q_{2m}+ q_{2l} q_{3m}+ q_{3l}
     q_{1m})\e_{jklm}\frac{q_{1i}}{\tau_1}-\e_{ijkl}
     \frac{1}{\tau_3} q_3^2 (q_1-q_2)_l
     +\mbox{cyclic}] \ ,
\err
where we used momentum conservation.  As the integral
is finite, we have set $d=4$.  Comparing with the supergravity 
result \eqn{further} we find that they agree.  The normalizations of the 
correlators match if 
\brr   
\frac{4}{3}  \pi^2 C_4 N^2= {k\over 4} \pi^8 \lambda^3 \ , 
\err 
i.e. if $k=-8N^2/3$.  This completes the matching of the two- and 
three-point correlators in the $\mbox{AdS}$/CFT correspondence.    

\section{Conclusions}
\setcounter{equation}{0}

In this work we have explicitly computed the two- and three-point correlators of the
$SU(4)$ $R$-current in $\cN=4$ super-Yang-Mills.   The correlators match precisely at
one-loop with the $\mbox{AdS}$ supergravity  correlation functions according to the
prescription of \cite{ew,gkp}.  The matching  at one-loop is clear for the anomalous
part of the correlator due to the  Adler-Bardeen theorem; the matching of the
vector-like contribution, however,  at exactly one-loop implies the vanishing of all
higher loop effects within  the $\mbox{AdS}$/CFT correspondence.  The latter points to the
existence of non-renormalization theorems in the superconformal theory.  In
\cite{freedman} this  was pointed out based on a different technique of computing the
$R$-correlators.   Keeping the results as integral expressions shows that the
vector-like portion of  the correlator is of the same form as the anomalous part.  The
three-point  function has highly constrained kinematics and its Schwinger
parameterization contains an exponential function only of the external momentum inner
products $q_i^2$.  This allows one to match the integrands in a rather straightforward
manner; in the super-Yang-Mills side  this dependence comes from the loop integration,
and in the $\mbox{AdS}$ side it arises from the boundary/bulk kernel.     
  
The off-shell generating functions for the $R$-symmetry correlators are difficult to
compute in pure field theory at higher-point, and our technique rather relies on
mapping the integral expressions to the appropriate supergravity results in  the
$\mbox{AdS}$/CFT  correspondence.  Our method of testing the conjectured $\mbox{AdS}$/CFT
correspondence is suitable for generalizing to further matching tests of higher-point
functions without the technical problems associated with the loop integrations.  The 
four-point function is interesting for several reasons, but unlike the  three-point
function its form  is not fixed by superconformal invariance.  The kinematical
constraints are  weaker at the four-point level and thus matching it in the AdS/CFT
correspondence is a more non-trivial test of the  conjecture.  However, there is no
obvious reason to believe that the correspondence of the higher-point functions holds
again  at one-loop.  Furthermore, the same analysis performed here may  clearly be
performed to relate other AdS three-point functions in the correspondence  besides the
$R$-symmetry currents to gauge theory correlators.   

\vspace{1cm} 
\noindent {\bf Acknowledgements}   
\vskip .23in
 
\noindent We thank I.\ Chepelev, R.\ Kallosh, M.\ Ro\v cek, N.\ Seiberg, W.\ 
Siegel, and P.\ van Nieuwenhuizen for discussions. We would also like to thank D. Freedman for drawing our attention to an error in an earlier version of this article.  This work was supported in part by NSF grant  No.~PHY 9722101.

\end{document}